\documentstyle[12pt]{article}
\newlength{\abstractwidth}
\renewcommand{\thefootnote}{\fnsymbol{footnote}}
\renewcommand{\thanks}[1]{\footnote{#1}} 
\newcommand{\starttext}{\setcounter{footnote}{0}
\renewcommand{\thefootnote}{\arabic{footnote}}}

\def \be {\begin{equation}}
\def \ee {\end{equation}}
\def \bea {\begin{eqnarray}}
\def \eea {\end{eqnarray}}

\def \a {\alpha}
\def \b {\beta}
\def \g {\gamma}

\begin{document}
\begin{titlepage}
\bigskip
\bigskip\bigskip\bigskip\bigskip
\bigskip \bigskip
\centerline{\Large \bf {Holographic Cosmic Quintessence}}
\bigskip
\centerline {\Large \bf  {on}}
\bigskip
\centerline{\Large \bf {Dilatonic Brane World }}
\bigskip\bigskip
\bigskip\bigskip
\centerline{Bin Chen$^{1}$\footnote{chenb@ictp.trieste.it}  and Feng-Li Lin
$^{2}$\footnote{linfl@phya.snu.ac.kr}}
\bigskip
\centerline{\it ${}^1$High Energy Section} \centerline{\it the Abdus Salam
ICTP} \centerline{\it Strada Costiera, 11} \centerline{\it 34014 Trieste,
Italy}
\bigskip
\centerline{\it ${}^2$School of Physics \& CTP} \centerline{\it Seoul National
University} \centerline{\it Seoul 151-742, Korea }
\bigskip\bigskip


\begin{abstract}
\medskip
\noindent  Recently quintessence is proposed to explain the observation data of
supernova indicating a time-varying cosmological constant and accelerating
universe. Inspired by this and its mysterious origin, we look for the
possibility of quintessence as the holographic dark matters dominated in the
late time in the brane world scenarios. We consider both the cases of static
and moving brane in a dilaton gravity background. For the static brane we use
the Hamilton-Jacobi method motivated by holographic renormalization group to
study the intrinsic FRW cosmology on the brane and find out the constraint on
the bulk potential for the quintessence. This constraint requires a negative
slowly varying bulk potential which implies an anti-de Sitter-like bulk
geometry and could be possibly realized from the higher dimensional
supergravities or string theory. We find the similar constraint for the moving
brane cases and that the quintessence on it has the effect as a mildly
time-varying Newton constant.
\end{abstract}
\end{titlepage}

\starttext \baselineskip=18pt
\setcounter{footnote}{0}


\setcounter{equation}{0}
\section{Introduction}

   Cosmic quintessence \cite{Stein} is proposed recently to be an alternative
to understand the astronomical data of supernova \cite{nova}(see also
\cite{summary} for a summary of the data constraints.) indicating an
everlasting accelerating universe, which can also be interpretated as the
effect of a small slowly time-varying cosmological constant \cite{Stein}.
Recently there are discussions on the difficulty about the string theory in a
spacetime with future cosmological horizon \cite{Witten,Susskind} which exists
both for de Sitter space and the universe dominated by cosmic quintessence at
late time, and hinders the notion of S-matrices in the perturbative string or
quantum field theory. However, from the cosmological point of view, the
quintessence proposal has milder cosmological constant problem without severe
fine tuning since it does not require a zero cosmological constant at all time,
instead it models a slowly decaying cosmological constant and its effect at the
present moment will fit the observational data by properly tuning its energy
density.

  Let us start with the usual (3+1)-dim. Friedman-Robertson-Walker(FRW) metric
\be
\label{mFRW}
ds^2_{FRW}=-d\tau^2+a^2(\tau)d\Sigma_k^2\;,
\ee
and the Einstein equations are reduced to the FRW equations
\bea
\label{FRW0}
H^2&=&{-k\over a^2}+\sum_i{\rho_i \over 3} ({a_0 \over a})^{\kappa_i}\;,\\
6{\ddot{a}\over a}&=&\sum_i(2-\kappa_i)\rho_i({a_0\over a})^{\kappa_i}\;.
\eea
Note that the second FRW equation is independent of $k$ where $k=-1,0,1$
corresponds to open, flat and closed universe with the corresponding
hypersphere $\Sigma_k$; $a(\tau)$ is the cosmological scale factor with $a_0$
parameterizing the initial size of the universe; $H=\dot{a}/a$ is the Hubble
parameter with $\dot{a}\equiv {d a\over d \tau}$. In the above we have assumed
the universe contains only the perfect fluids of pressure $p_i$ and the energy
density $\rho_i=\rho_{0i} (a_0/a)^{\kappa_i}$, and obeying the equation of
state $p_i=\omega_i \rho_i$ with
\be
\kappa_i=3(1+\omega_i)\;.
\ee

Usually one could have different types of matters present in the universe, for
example, a CFT matter or photon has $\omega=1/3$(or $\kappa=4$), and a
cosmological constant corresponds to $\omega=-1$(or $\kappa=0$), and the usual
matters have $\omega=0$($\kappa=3$). In the late time or for the large scale
factor, the dynamics of the FRW cosmology is dominated by the component with
small $\kappa>0$.

In this paper, the dominated matter in the late time is assumed to be the
quintessence in accord with the observation data in \cite{nova} giving the
following constraint on $\omega$
\be
-1<\omega_{obs}<{-2\over3}\ \qquad or \qquad 0<\kappa<1\;,
\ee
which indicates the universe is accelerating at the present stage according to
the 2nd FRW equations for both flat, open and closed universes. In general an
accelerating universe requires
\be
\label{quint}
-1<\omega<{-1\over 3}\ \qquad or \qquad 0<\kappa<2\;,
\ee
which is driven by some matters coined as quintessence dominated at the late
time of cosmic evolution \cite{Stein}. As emphasized in \cite{Susskind}, an
observer in an accelerating universe will see a future horizon.

   In this paper we will investigate the possibility of having the quintessence as
the holographic imprint of the brane-world scenarios.  Since the quintessence
can be in general modelled as a slowly-rolling scalar which does not exist in
the usual Randall-Sundrum(RS) type brane-world scenarios \cite{RS} with a brane
embedded in the pure AdS space. However, in the dilatonic gravity it is natural
to have a Liouville type potential, that is, in the exponential form. As one
follows the AdS/CFT correspondence \cite{adscft}, the bulk gravity will induce
holographic matters with similar on-shell Liouville type potential on the
brane, which can be interpretated as the holographic dark energy from the point
of view of brane cosmology; and for some specific dilaton gravity one may
obtain the holographic cosmic quintessence.

     Our answer to this investigation is positive. In the next section 2 and 3
we will review the general holographic principle in dilaton gravity which is
intimately related to the holographic renormalization group \cite{BVV} and
induces intrinsic geometry on the brane. Based upon this intrinsic brane
gravity we will discuss the possibility of having the holographic cosmic
quintessence on the static brane. In section 4 and 5 we will turn to the more
general bulk dilatonic background and discuss the holographic quintessence on a
moving brane based upon the FRW cosmology derived from the extrinsic geometry
of the brane, the Israel junction condition. It turns out that the quintessence
on the moving brane has the similar effect as a mildly time-varying Newton
constant. In the last section we conclude with some remarks and discussions.

\setcounter{equation}{0}
\section{Dilatonic Brane Gravity from Holography}
    The Randall-Sundrum model \cite{RS} of branes embedded in a bulk
(anti-de Sitter) spacetime can be thought as a variant of the AdS/CFT
correspondence \cite{adscft}, that is, the dual CFT is coupled to the brane
gravity as the induced holographic dark matters \cite{Gubser,Hawking,Nunez}.
After integrating out the hidden CFT matters we will obtain an effective
gravity theory in which we can add the local matters like the Standard model on
the curved brane. This consideration of holography can be generalized to the
bulk dilatonic gravity with the action
\be
\label{action}
S=\int_{\cal M} d^{n+2}x \sqrt{-g} ({1\over2}R-{1\over 2} g^{MN} \partial_{M}
\phi
\partial_N \phi -V(\phi))\;,
\ee
where $V(\phi)$ is the dilaton potential which usually have its origin in
higher dimensional supergravity or superstring theory. After integrating out
the dual holographic field theory on the brane, one will obtain an effective
dilaton gravity theory on the brane with the following action
\be
\label{ba}
S_{\Sigma}= \int_{\Sigma}d^{n+1}x \sqrt{-\gamma}\; (Z(\phi){\cal
R}-{1\over2}M(\phi)\gamma^{\mu\nu}\partial_{\mu}\phi\partial_{\nu}\phi-U(\phi))
+\Gamma[\gamma,\phi]\;,
\ee
where ${\cal R}$ is the scalar curvature with respect to the boundary metric
$\gamma_{\mu\nu}$. The UV part $\Gamma$ contains the higher-derivative and
non-local terms and thus can be neglected in the low energy regime\footnote{The
trace of the variation of $\Gamma[\gamma,\phi]$ with respect to $\g_{\mu\nu}$
is the conformal anomaly which is argued in \cite{Odintsov1,Hawking} to drive
the inflation of the brane universe, and is thus relevant in the early universe
in accord with its UV nature.}.

The boundary functionals $Z(\phi)$, $M(\phi)$ and $U(\phi)$ on the brane
worldvolume $\Sigma$ are not arbitrary but are determined or induced by the
bulk gravity theory and the geometry, and can be viewed as the holographic dark
energy on the brane( or domain wall). In the bulk geometry of the
Randall-Sundrum type with the metric
\be
\label{metric1}
g_{MN}dx^{M}dx^{N}=dr^2+ds^2_{brane}\equiv dr^2+\Lambda^2(r)\bar{\gamma}_{\mu
\nu}(x^{\mu})dx^{\mu}dx^{\nu}\;,
\ee
where $\bar{\gamma}_{\mu\nu}$ depends only on the brane coordinates not on $r$,
and the on-shell bulk and boundary potentials are related by the following
"superpotential-like" relation \cite{Gubser,superpotential}
\be
\label{bubn}
{1\over 2}({\partial U \over \partial \phi})^2-{2\over3}U^2=V(\phi)\;.
\ee

A more general and systematic discussions based upon the Hamilton-Jacobi
formalism about the holographic relations was given in \cite{BVV} (see also
\cite{Shiromizu} for the recent attempt in the cosmology context). This
formalism is a generalization of AdS/CFT correspondence, and the dilaton is
interpretated as the effective coupling of the holographic renormalization
group flow in the boundary theory with its beta function defined by
\be
\beta=\Lambda{\partial \phi \over \Lambda}\;,
\ee
where $\Lambda$ is the overall scale factor in (\ref{metric1}) and is
interpretated as the holographic energy scale in the dual theory.

   On the other hand, one can determine the intrinsic brane gravity
described by $S_{\Sigma}$ from the Hamilton-Jacobi equation given in \cite{BVV}
which is
\be
{\cal H}[\pi_{\sigma \nu}= {1\over \sqrt{det\;\g}} {\delta S_{\Sigma} \over
\delta \g^{\sigma \nu}},\Pi = {1\over \sqrt{det\;\g}} {\delta S_{\Sigma} \over
\delta \phi}]=0\;,
\ee
where ${\cal H}$ is the Hamilton density constructed from the action
(\ref{action}) in the standard ADM formalism. The $\pi_{\mu\nu}$ and $\Pi$ are
the canonical conjugate momenta with respect to $\gamma_{\mu\nu}$ and $\phi$.
The boundary action $S_{\Sigma}$ is then identified as the Hamilton-Jacobi
functional as indicated above. From the Hamilton-Jacobi equation, we obtain the
following independent holographic relations \cite{BVV,linwu}\footnote{Our
convention taken here is different from the one in \cite{BVV,linwu} with an
additional factor $1/2$ in front of the bulk scalar curvature term so that the
coefficients in the following holographic relations change accordingly.}
\bea
{-4\over3}UZ+2U^{'}Z^{'}=1\;,
\\
\beta=12{Z^{'}\over M}\;,
\\
-4M-24Z^{''}+\beta M^{'}={6\over U}\;,
\eea
where $'$ means the derivative with respect to $\phi$. Moreover, the beta
function can be determined from inverting the defining equations of the
canonical momenta, and it turns out to be
\be
\label{Ub}
\beta=-{6U^{'}\over U}\;.
\ee

   From the set of the holographic relations we can determined the brane
gravity theory with a given boundary dilaton potential. It is easy to see that
there is no nontrivial solution for constant $M$ and $Z$, so it implies that
the brane gravity usually has a nontrivial dilaton potential.  Since we are
interested in the possibility of the holographic quintessence which can be
modelled with a power-law or exponential form dilaton potential, we try the
ansatz for such kind of the potential form.  We find that there is no simple
solution for the power-law potential, while we have the solution for the
Liouville type as the following
\bea
\label{U}
U&=&e^{b(\phi-\phi_0)}\;,
\\
\label{MZ}
M&=&2Z=({-6\over 4+6b^2}){1\over U}\;,
\eea
for arbitrary constants $b$ and $\phi_0$\footnote{
In fact, the general solutions for $Z$ and $M$ are
\bea
Z&=&c_0e^{2\phi \over 3b}-{3 \over 4+6b^2}e^{-b\phi} \nonumber \\
M&=&-{4c_0 \over 3b^2}e^{2\phi \over 3b}-{6 \over 4+6b^2}
e^{-b(\phi-\phi_0)}\nonumber
\eea
where $c_0$ is an integration constant. Since the $c_0$ terms have the opposite
running behavior from the other terms and will destroy the UV/IR relation in
the holographic RG scheme, thus we should put it to zero for consistency.}.
Moreover, the bulk and boundary potential is related by (\ref{bubn}) as
\be
\label{HoloV}
V=({3b^2-4 \over 6})\; U^2 \;.
\ee
Note that for $3b^2=4$, $V=0$ which corresponds to the self-tuning brane model
of \cite{selftune}.  In the next section we will study the FRW cosmology of the
brane gravity $S_{\Sigma}$ to check the possibility of the holographic cosmic
quintessence for the static brane.

\setcounter{equation}{0}
\section{Quintessence on the Holographic Brane}

Before going to the FRW cosmology of the brane gravity described by
$S_{\Sigma}$ we first recall some facts about the FRW cosmology of the
quintessence modelled by the following (3+1)-dim. scalar Lagrangian
\cite{Peebles}
\be
\label{L}
{\it L}=\sqrt{\gamma}({1\over2}(\partial\Phi)^2+W(\Phi))\;,
\ee
With respect to the FRW metric (\ref{mFRW}), the energy density
$\rho$ and the pressure $p$ take the form
\be
\label{rho}
\rho={1\over2}{\dot\Phi}^2+W(\Phi)\;, \qquad
p={1\over2}{\dot\Phi}^2-W(\Phi)\;,
\ee
and the equation of state is
\be
\label{ep}
\omega\equiv{p\over\rho}={\dot{\Phi}^2-2W(\Phi) \over
\dot{\Phi}^2+2W(\Phi)}\;.
\ee
Note that when $\dot{\Phi}^2<W(\Phi)$ in the late time, then $\omega<-1/3$(or
$\kappa<2$) giving the quintessence. In the case of $k=0$, it has been shown in
\cite{Peebles} that the only stable attractive fixed point of the solution
space for $\kappa=3(1+\omega)<2$ is that $\Phi$ scales as
\be
\label{run}
a{\partial \Phi \over \partial a}=\sqrt{\kappa}\;,
\ee
and the scalar energy density red-shifts as
\be
\rho=\rho_0({a_0\over a})^{\kappa}
\ee
as usual. Note that the scalar-field equation of motion implies the
conservation of the scalar-field stress tensor. At this fixed point it is easy
to see that
\be
W(\Phi)=W_0 e^{-\sqrt{\kappa}\Phi}\;.
\ee

 The above indicates that a scalar field with Liouville type potential
could be the quintessence within the usual Einstein gravity. However, the brane
gravity in the last section is the gravity not in the Einstein frame but in the
stringy frame with nontrivial $Z(\phi)$ and $M(\phi)$ in the action
$S_{\Sigma}$.  In order to study the usual FRW cosmology, we need to change to
the Einstein frame by the following Weyl transformation\footnote{There is an
issue about the equivalence of the different conformal frames as discussed in
\cite{Battye,conformal,Odintsov2}, we adopt the point of view that the standard
FRW cosmology should be considered in the Einstein frame, we also find that the
boundary gravity in the stringy frame (\ref{ba}) will lead to negative energy
density for the FRW cosmology which agrees with the observation in
\cite{Battye}.}
\be
\hat{\gamma}_{\mu \nu}=Z(\phi)\gamma_{\mu\nu}\;,
\ee
and with the $Z$, $M$ and $U$ of (\ref{MZ}) and (\ref{U}) the new action in the
Einstein frame\footnote{In this section we adopt the convention in
\cite{Peebles} such that there is no $1/2$ factor in front of the
Einstein-Hilbert term in (\ref{ac}).} is
\be
\label{ac}
S_E=\int d^{3+1}x\sqrt{-\hat{\gamma}}[R-{1\over 2}(\hat{\partial}\Phi)^2
-({4+6b^2\over3 })^2 e^{{3b\over \sqrt{2+3b^2}}(\Phi-\Phi_0)}]\;,
\ee
where we have defined
\be
\Phi=\sqrt{2+3b^2}\phi\;.
\ee
so that the scalar kinetic term is in the canonical form. Note that the
resulting potential in the Einstein frame is positive definite so that the
positive energy density of the holographic matters is ensured.

   In the Einstein frame the brane gravity has the scalar-field Lagrangian
$L$ of (\ref{L}) with a specific potential $W(\Phi)$ by identifying
\be
\sqrt{\kappa}={-3b\over \sqrt{2+3b^2}}\;,
\ee
which then requires $b<0$\footnote{The condition $b<0$ is consistent with the
usual runaway behavior of the string dilaton so that the string coupling
$g_s=e^{-b\phi}\rightarrow 0$ as $\phi \rightarrow \infty$.} and the
quintessence condition $0<\kappa<2$ requires
\be
0<b^2<{4\over 3}\;.
\ee
Note that the marginal point for quintessence $b^2=4/3$ is the self-tuning
brane condition for zero cosmological constant\footnote{For $b^2=4/3$ the
potential in the Einstein frame is nonzero, this may indicate the inequivalence
between the intrinsic holographic gravity and the extrinsic approach based upon
some generalization of the method developed by \cite{BDL,Battye}; however, it
needs further investigations to clarify this point.} . From the holographic
relation (\ref{HoloV}), we conclude that a bulk potential
$V(\phi)=V_0e^{2b\phi}<0$ with $0<b^2<4/3$ will induce holographic quintessence
on the brane cosmology. We want to emphasize the fact $V_0<0$ for $0<b^2<4/3$
because this leads to an anti-de Sitter-like bulk geometry which is far more
easy to be realized in the supergravity than the de-Sitter-like one
\cite{Witten}. It is interesting to look for the origin of $V(\phi)$ in the
higher dimensional supergravities or string theory.

    There are further complications by adding the standard matter fields besides the
above Liouville scalar to cook up the more realistic cosmological model, the
related issues and results have been discussed in \cite{Peebles,Kolda} and can
be directly carried to the holographic brane scenarios discussed here.

\setcounter{equation}{0}
\section{FRW Cosmology on the Moving Brane}

 In the last section we have considered the brane world scenarios as the
generalization of AdS/CFT correspondence, by which the intrinsic geometry on
the static brane is induced by the dual QFT on the brane. Although it is very
intriguing to have covariant intrinsic gravity on the brane, it is difficult to
generalize such program to a moving brane since the energy scale on the brane
keeps changing all the time, which looks bizarre from the holographic RG point of
view. Instead an extrinsic approach is more appropriate in which the brane
cosmology is derived from the extrinsic geometry of the brane by using the
Israel patching condition \cite{Kraus}. However, from the general
holographic principle point of view, we should expect the equivalence of the
intrinsic and extrinsic approaches, which remains to be explored. Moreover,
the FRW cosmology for a moving brane is also known as the mirage cosmology from
the point of view of the effective theory on the brane \cite{mirage}, that is,
the holographic image of the bulk gravity serves as the dark energy which
drives the evolution of the brane universe.

To derive the brane FRW equations for a brane moving along the radial
direction, we can add the following Gibbons-Hawking boundary action to
(\ref{action})
\be
S_{GH}=-\int_{\Sigma}d^{n+1}x \sqrt{-\gamma}\;K\;,
\ee
where $K$ is the extrinsic curvature.
   The brane FRW equations can be obtained from the following Israel junction
condition with brane situated at $r=R$ \cite{Chamblin,Kraus}
\be
\label{junc}
\Delta K_{\mu\nu}=K_{\mu\nu}(r=R_+)-K_{\mu\nu}(r=R_-)=-(T_{\mu\nu}-{1\over n}
T^{\gamma}_{\gamma} \gamma_{\mu\nu})\;,
\ee
which specifies the patching condition of the bulk geometry on the two sides of
the brane with the boundary energy-momentum tensor $T^{\mu\nu}\equiv
{2\over\sqrt{-\gamma}}{\delta S_{\Sigma} \over \delta \gamma_{\mu\nu}}$. As
remarked at beginning of this section, only the effective tension term of the
brane, $U(\phi)$ of $S_{\Sigma}$ is kept since we are taking an extrinsic
approach, of which there is no essential information about the intrinsic
geometry on the brane. Moreover, by assuming the bulk spacetime is symmetric
under the reflection with respect to the brane, (\ref{junction}) becomes
\be
\label{junction}
K_{ij}=-{1\over 2n} \gamma_{ij} U(\phi)\;.
\ee

   Rewriting the junction condition (\ref{junction}) into the FRW equations on the
brane has been done in \cite{Gubser,Visser,Verlinde} for the AdS gravity of
which $U(\phi)$ is a constant, especially a moving brane embedded in a Anti-de
Sitter-Schwarzschild space results in a brane FRW cosmology with the CFT
matters \cite{Gubser,Verlinde}. Here we will generalize it to the dilatonic
gravity with nontrivial $U(\phi)$ for the following bulk geometry
\be
g_{MN}dz^{M}dz^{N}=-f(r)dt^2+g(r)dr^2+h(r)\delta_{ij}dx^{i}dx^{j} \equiv
g(r)dr^2+\gamma_{\mu\nu}dx^{\mu}dx^{\nu}\;,
\ee
with $i,j$ run from $1$ to $n$.

For the simplicity, we now restrict to the $n=3$ case, and consider a moving
brane with the trajectory described by $R(\tau)$ where $\tau$ is the proper
time defined by
\be
-d\tau^2=-f(R)dt^2+g(R)dR^2\;,
\ee
then the brane metric takes the FRW form of (\ref{mFRW}) with
$a^2(\tau)=h(R(\tau))$. Once the trajectory of the moving brane is assumed with
respect to the proper time, we can calculate the extrinsic curvature
$K_{\mu\nu}\equiv{1\over 2} n^M\partial_M\gamma_{\mu\nu}$ with the unit normal
$n^M$ defined with respect to the velocity vector
$u^{M}=(\dot{t},\dot{R},0,0,0)$, where dot denotes the derivative withe respect
to $\tau$. The junction condition (\ref{junction}) then turns out to have the
form of FRW equation for the cosmology on the moving brane as
\be
\label{FRW}
({\dot{a} \over a})^2={-1\over 4 g}({h^{'}\over h})^2+({U\over6})^2\;.
\ee

\bigskip

   To get familiar with the above FRW equation, we consider the simplest case
for the bulk to be the pure black hole background without dilaton, we have
$f(r)=g^{-1}(r)=k-V_0r^2/3-M/r$ with $k=-1,0,1$, $h(r)=r^2$ and $U=U_0,V=V_0$
with $U_0,V_0$ constants, then the FRW equation (\ref{FRW}) becomes
\be
\label{FRW1}
({\dot{a}\over a})^2={-k\over a^2}+{M\over a^4}+{V_0\over3}+({U_0\over 6})^2\;.
\ee
It is clear that this corresponds to the (3+1)-dim. radiation dominated FRW
cosmology with a cosmological constant proportional to ${V_0\over3}+({U_0\over
6})^2$ which can be fine tuned to zero only for bulk anti-de Sitter space with
$V_0<0$. This kind of fine-tuning is the brane version of cosmological constant
problem.

  Another interesting example is the dilatonic background for the self-tuning (static)
brane proposed in \cite{ADKS,selftune} to solve the brane version of the
cosmological constant problem, in which $V_0=0$ but $U=U_0 e^{-\sqrt{{4\over
3}}\phi}$, and the metric is\footnote{Since we have a different normalization
of the scalar kinetic term from the one in \cite{ADKS,selftune}, the
corresponding factors in the $A$ and $\phi$ change accordingly.}
\be
ds^2=dr^2+ e^{2A(r)}dx^2_{brane}\;,
\ee
where $A(r)$ is related to the dilaton profile which is
\be
\phi(r)=2\sqrt{3}A(r)=\sqrt{{3\over4}}\log|{4r\over3}+c_1|+c_0\;,
\ee
with $c_{0,1}$ the integration constants which can be exploited to tune the
brane cosmological constant \cite{ADKS,selftune}. It is easy to see that the
bulk geometry has a naked singularity at finite $r$ although it has been argued
that this kind of singularity are harmless according to the criterions given in
\cite{singu,kimd}.

Instead of considering the static brane, we now embed a moving brane in such a
background, the brane cosmology is then described by the following FRW
equations
\be
\label{self}
({\dot{a}\over a})^2=({-e^{{4c_0\over \sqrt{3}}}\over 9}+({U_0\over
6})^2){1\over a^8}\;.
\ee
 Note that the scale behavior of
the holographic matter on the RHS of the above FRW equation is neither of CFT
nor of the quintessence. In some sense the above FRW cosmology has zero
cosmological constant since the holographic matter is highly damped for the
large scale factor and then leave nothing on the RHS of the FRW equation. This
fact coincides with the claim of self-tuning cosmological constant for the
static brane case \cite{ADKS,selftune}.

   Moreover, if the holographic matter in (\ref{self}) has positive energy by
proper choice of the parameters $C_0$ and $U_0$, then the brane is inflating
and moving away from the naked singularity which thus will not cause any
physical problem in the history of the evolution of the brane universe. In this
sense, the naked singularity is harmless. On the other hand, if the holographic
matter has negative energy, then the brane world is moving toward the
singularity and will finally crash on it even our universe is flat. The
analogue of a naked singularity in the brane world scenarios and the big
bang( crunch) singularity has been exploited in \cite{kimd} to give some
criterion for the allowing naked singularities which appear generically in many
bulk geometry, and this kind of criterion is argued to be equivalent to the one
in \cite{singu}.

   Despite that the self-tuning brane model provide a possible solution
to the cosmological constant problem, from the point of view of the FRW
cosmology, the holographic matters induced on the brane do not yield the
quintessence. In the next section we will see that some more general dilatonic
background does induce holographic quintessence.

\setcounter{equation}{0}
\section{Holographic Quintessence on the Moving Brane}

   There are three types of dilatonic backgrounds \cite{Chamblin}(see also \cite{Cai1}.)
which solve the bulk equations of motion of (\ref{action}) with the bulk and
boundary dilaton potentials taken the following form
\be
V=V_0e^{\b \phi}\;, \qquad U=U_0 e^{\a \phi}\;.
\ee
   The cosmological behaviors on the brane in these backgrounds
have been discussed in details in \cite{Chamblin}, in the following we
recapitulate their results but focus on the possible existence of the
quintessence.

   The 1st type background is just the one given in the last section with
$\a=\b=0$ which gives the radiation dominated FRW cosmology. We then look into
the 2nd type dilatonic background with $\a=\b/2$,$k=0$ and
\be
f(r)=g^{-1}(r)=(1+\delta^2)^2r^{2\over
1+\delta^2}(-2Mr^{-{4-\delta^2\over1+\delta^2}}-{2V_0e^{2\delta\phi_0}\over
3(4-\delta^2)})\;,
\ee
\be
h(r)=r^{2\over 1+\delta^2}\;, \qquad \phi(r)=\sqrt{3}(\phi_0-{\delta\over
1+\delta^2} \log r)\;,
\ee
where $M$ and $\phi_0$ are integration constants and $\delta={\sqrt{3}\b \over
2}=\sqrt{3}\a$.

The FRW equation (\ref{FRW}) of the brane in this background becomes
\be
\label{FRW2}
({\dot{a} \over a})^2= {2M\over a^{(4+\delta^2)}}+{\hat{V}_0 \over
a^{2\delta^2}}\;.
\ee
where $\hat{V}_0\equiv ({2V_0\over 3(4-\delta^2)}+{U^2_0\over
36})e^{2\delta\phi_0}$. In the late time the second term dominates and is
quintessential if
\be
\delta^2<{1}\;.
\ee
To ensure the positive energy condition for the all time, it requires
$M,\hat{V}_0>0$ which corresponds to a bulk dilatonic black hole background for
$V_0<0$ \cite{Cai1,Chamblin}. Moreover, the factor $e^{2\delta\phi_0}$ implies
no need for fine-tuning for the energy density of the quintessence.

  Similarly, the brane FRW cosmology in the 3rd type dilatonic background is
\be
({\dot{a}\over a})^2=2\delta^4M({a_0\over a})^{4+{1\over \delta^2}}+
{2\delta^4V_0e^{2\delta\phi_0}\over 3(1+2\delta^2)}({a_0\over a})^2
+({U_0e^{\phi_0\over \delta} \over 6})^2({a_0\over a})^{2\over \delta^2}\;.
\ee
and the background geometry is given by
\be
f(r)=g^{-1}(r)=(1+\delta^2)^2r^{2\over
1+\delta^2}(-2Mr^{-{1+2\delta^2\over1+\delta^2}}-{2V_0e^{2\delta\phi_0}\over
3(1+2\delta^2)})\;,
\ee
\be
h(r)=a_0^2r^{2\delta^2 \over 1+\delta^2}\;,\qquad
\phi(r)=\sqrt{3}(\phi_0-{\delta \over 1+\delta^2} \log r)\;,
\ee
and $\delta={\sqrt{3}\b \over 2}={1\over \sqrt{3}\a}$, $a_0
=|{V_0e^{2\delta\phi_0}(1-\delta^2)\over 3}|^{-1/2}$.

It is then easy to see that in the late time the quintessence requires
\be
\delta^2>1\;.
\ee
For the bulk geometry to be a black hole type, it requires $V_0<0, M>0$
\cite{Cai1,Chamblin}, then the positive energy condition may be violated.

   To explore a little bit on the nature of the above quintessence, we can
add the local matter on the brane. As usual, we focus on the perfect fluid
matter with energy-momentum tensor
\be
T^{(m)}_{\mu\nu}=\rho^{(m)} u_\mu u_\nu+p^{(m)}\;(\gamma_{\mu\nu}+u_\mu u_\nu)
\ee
where $\gamma_{\mu\nu}=g_{\mu\nu}-n_\mu n_\nu$. Then the FRW equation from the
Israel junction condition changes to
\be
(\frac{\dot{a}}{a})^2=-\frac{1}{4g}(\frac{h^\prime}{h})^2+(\frac{\rho^{(m)}+U}{6})^2
\ee

For the 1st type bulk background $U$ is constant, so we have the well-known
$(\rho^{(m)})^2$ term different from the conventional FRW cosmology \cite{BDL}.
On the other hand, for the 2nd and 3rd type quintessential brane, $U$ is not a
constant and will dress its scale dependence on the $\rho^{(m)}$ and
$(\rho^{(m)})^2$. If we restore the Newton constant dependence in the FRW
equation, that is,
\be
H^2={8\pi G_N \over 3}\rho\;,
\ee
we can absorb the dressing scale dependence of the $\rho^{(m)}$ term into the
Newton constant $G_N$ so that the matter density $\rho^{(m)}$ scales as
required by the energy conservation and equation of state. This implies that
the quintessence on the moving brane has the nature of a {\it mildly}
time-varying Newton constant since the dressing scale dependence due to the
quintessence is very slowly varying. Similar conclusion about the time-varying
Newton constant for a moving brane of zero cosmological constant is also
discussed in \cite{Cai2}. We also expect the similar situation for the static
brane case. Moreover, these dressing will red-shift the matter energy.

\setcounter{equation}{0}
\section{Conclusion}
Dilaton definitely has its deep origin in string theory and supergravity. In
this paper we show that some dilaton potentials will induce the holographic
quintessences on the brane, which provides a natural candidate to conform to the
cosmological observational data in the brane world scenarios.  By the way, it is
interesting to compare our result with the efforts in getting the quintessences
from some dilaton potentials in supersymmetric theory of supergravity origin. In
\cite{Susskind} it has been shown that the quintessential behavior is
incompatible with the positive energy condition in the case of the dilaton
potential in the suspersymmetric theory.  However, as shown in this paper,
the holographic principle shows that the boundary
potential is related to the bulk potential by some inverse super-potential like
relation, and there is no inconsistency between
quintessence and positive energy condition.

We also see that the intrinsic brane gravity based upon the Hamilton-Jacobi
formalism of holographic RG approach provide a useful tool in studying the
brane cosmology. But it is speculative to claim the equivalence between this
intrinsic approach and the extrinsic ones \cite{BDL} in generalizing to the
dilaton gravity \cite{Cline,Battye}, which requires more investigations. It
deserves more study on the phenomenological aspects of the above quintessential
brane world scenarios. Especially the quintessence on the moving brane has the
effect as the time-varying Newton constant which may lead to some novel effect
beyond the standard FRW cosmology. Moreover, it is also interesting to look
into the interplay between the sectors of holographic and normal matter to find
a way out of the eternal acceleration of the quintessential universe in the
brane world scenarios.

\vskip 1cm
\centerline{\bf Acknowledgements}
\vskip 1cm

 BC would like to thank  A. Mazumdar for valuable comments and reading the manuscript.
FLL is supported by BK-21 Initiative in Physics (SNU - Project 2), he is also
grateful to the hospitality of ICTP at Trieste and KIAS at Seoul where this
work was initialized during his visit, and also thanks Hyung Do Kim for very
helpful and stimulating discussions.

\end{document}